# X-ray Free Electron Laser Studies of Electron and Phonon Dynamics of Graphene Adsorbed on Copper

Hirohito Ogasawara[1], Han Wang[1,7], Jörgen Gladh[1,2], Alessandro Gallo[1], Ralph Page[1,2], Johannes Voss[1], Alan Luntz[1]*, Elias Diesen[1,9], Frank Abild-Pedersen[1], Anders Nilsson[1,3], Markus Soldemo[1,2,3], Marc Zajac[2,8], Andrew Attar[1], Michelle E. Chen[2], Sang Wan Cho[4], Abhishek Katoch[4], Ki-Jeong Kim[5], Kyung Hwan Kim[6], Minseok Kim[5], Soonnam Kwon[5], Sang Han Park[5], Henrique Ribeiro[1,2], Sami Sainio[1], Hsin-Yi Wang[3], Cheolhee Yang[6], Tony Heinz[1,2]

[1] SLAC National Accelerator Laboratory, 2575 Sand Hill Rd, Menlo Park, CA 94025, USA
[2] Department of Applied Physics, Stanford University, Stanford, California 94305, USA
[3] Department of Physics, Stockholm University, SE-10691 Stockholm, Sweden
[4] Department of Physics, Yonsei University, Wonju, Korea
[5] Pohang Accelerator Laboratory, Pohang, Korea
[6] Department of Chemistry, Pohang University of Science and Technology (POSTECH), Pohang, Korea
[7] School of Physical Science and Technology, ShanghaiTech University, Shanghai 201210, China
[8] X-Ray Science Division, Argonne National Laboratory, Lemont, IL 60439, USA

* acluntz@stanford.edu

**Abstract:**
We report optical pumping and X-ray absorption spectroscopy experiments at the PAL free electron laser that directly probe the electron dynamics of a graphene monolayer adsorbed on copper in the femtosecond regime. By analyzing the results with ab-initio theory we infer that the excitation of graphene is dominated by indirect excitation from hot electron-hole pairs created in the copper by the optical laser pulse. However, once the excitation is created in graphene, its decay follows a similar path as in many previous studies of graphene adsorbed on semiconductors, *i e.* rapid excitation of SCOPS (Strongly Coupled Optical Phonons) and eventual thermalization. It is likely that the lifetime of the hot electron-hole pairs in copper governs the lifetime of the electronic excitation of the graphene.

Fundamental dynamical processes of adsorbates on metal surfaces following laser excitation are often in the femtosecond (fs) regime and this has spawned the field of femtochemistry at surfaces. [1] These have historically involved fs laser optical pump- optical probe studies, with different optical nonlinear responses used to infer the dynamics, *e g.* two photon correlations, sum frequency generation, etc. While these studies can be quite sensitive to the nuclear dynamics, they are rather insensitive to the actual electron dynamics. In contrast, photoemission and X-ray spectroscopies directly measure electronic structure of the adsorbate-substrate system. For example, although UV photoemission measures the occupied valence bands of adsorbates on metals, both the metal and adsorbate can contribute to the photoemission so that there can be some ambiguity in the interpretation. However, X-ray spectroscopies such as X-ray absorption spectroscopy (XAS) and X-ray emission spectroscopies (XES) can also measure the electronic structure of the valence bands but in an element specific and symmetry selective way that completely isolates the electronic structure of the adsorbate from the metal. By combining such measurements with well-established theory one can also indirectly infer the adsorbate nuclear structure. [2] The emergence of X-ray free electron lasers (XFELs) over the past decade have opened up the possibility for optical pump -X-ray spectroscopy probe to directly measure the electron dynamics of the adsorbate and by combining with theory to infer its nuclear dynamics, both occurring in the fs regime. [3,4] For example, studies of the CO oxidation on Ru(0001) have allowed observation of a species highly excited close to the transition state appearing ~ 1 ps after initial optical excitation. [5] Similar experiments also allow the observation of short lived chemical transients in catalytic reactions, *e g*. the transition intermediate HCO during CO hydrogenation on Ru(0001) which has a lifetime of only ~ 1.5 ps. [6]

One phenomenon that is generally too fast for even the current generation of fs lasers and XFELs is to directly measure the lifetime of valence excited states of strongly bonded adsorbates on metal surfaces. This is because charge transfer between the adsorbate and metal surface is extremely fast, typically in the 1-3 fs time regime for strongly chemisorbed adsorbates and 10s of fs or longer for physisorbed species, and this limits the excited state lifetimes to the same time scale. [7] However, the dynamic consequences of valence excitation can sometimes still be observed, especially for physisorbed species with somewhat longer excited state lifetimes. Graphene adsorbed on copper is a well-known example of weak adsorbate bonding and has been well-studied structurally since CVD growth of graphene on copper is a commercial process for producing graphene. Because of the importance of graphene in emerging electronics and optoelectronics applications, there have been a large number of fs transient optical adsorption and photoemission studies of excited graphene's hot carrier dynamics, usually when free standing or supported on semiconductors such as SiC or insulator materials such as quartz. [8–12] These show that the initially created hot electron and hole pairs (e-h) in graphene at ½ $\hbar\omega_{optical}$ rapidly (~ 20 fs) form a hot quasi-thermal distribution of e-h pairs through interband scattering followed by energy transfer to a few strongly coupled optical phonons (SCOPS) on time scales of ~ 200 fs. The SCOPS act as a bottleneck for further e-h cooling which is then dominated by the energy transfer from the SCOPS to the remaining phonons of Gr occurs on ~ 1-3 ps time scales, with the rate likely dependent on both defect concentrations (supercollisions) and optical pump fluence. [12,13] The overall cooling of carriers following optical excitation is generally described by a three-temperature model (3T) with separate temperatures for the hot carriers, the SCOPS and the acoustic phonon bath. [14,15]

In this paper, we report measurements of the electronic structure of a weakly adsorbed graphene overlayer on a metallic copper substrate (Gr/Cu) following fs optical laser excitation using the PAL free electron X-ray laser as probe, [16,17] and with theory infer some of the fs resolved nuclear dynamics occurring after the optical pump. We utilize C K-edge X-ray absorption spectroscopy (XAS) to study the temporal evolution in the $\pi$ and $\pi^*$ valence bands of graphene following pulsed optical laser excitation at 400 nm. We find that the optical pump creates an immediate ~ 200- 300 fs transient decrease in population of the $\pi$ band some 1.7 eV below the Fermi energy and a related transient increase in the $\pi^*$ band just above the Fermi energy. We interpret this as evidence that the dominate electronic excitation arises from charge transfer processes from the dominantly excited copper substrate to the graphene. In addition, we utilize comparisons of the spectral changes to theory to argue that the longer temporal behavior of the XAS in the $\geq$ 300 fs regime describes the rapid evolution of the electronic excitation into SCOPS followed by a slower eventual thermalization of the SCOPS excitation into all phonon modes of graphene.

The optical pump- X-ray probe experiments involve following the electronic and structural relaxation in the Gr/Cu using time-resolved C 1s XAS measurements following the optical pump. Figure 1 shows a diagram of the experiment. The optical pump (400 nm, 3.1 eV photon energy, pulse width 100 fs) and x-ray probe (near the C 1s absorption edge, 281-290 eV photon energy, pulse width 50 fs) beams are incident collinearly on a flat Gr/Cu surface at a 20-degree grazing incidence geometry. The relative timing between the pump and probe beams is controlled with an optical delay line, with the spatial and temporal overlap (delay = 0) set by fluorescence of a thin Ce:YAG crystal. We estimate a temporal resolution of ~ 150 fs in the delay from the temporal fits to the spectral changes, largely limited by optical and X-ray beam spatial fluctuations. We record the C 1s XAS by Auger electron yield using a biased partial electron yield detector. The optical laser was p- polarized to induce strong optical adsorption in the Cu. The X-ray probe beam was also p-polarized, in which XAS dominantly sees the excitations involving $\pi$ or $\pi^*$ valence bands of Gr/Cu in the grazing incidence geometry. Transient response in the C 1s XAS after the optical pump can arise due to either occupation changes of $\pi$- and $\pi^*$- valence bands or structural modifications of the Gr. Considerably more experimental details are given in the supplementary material (SM).

Figure 2a (top) shows the XAS of Gr/Cu in the region of the $\pi^*$ resonance. The spectra are dominated by the excitation from a C1s to the $\pi^*$ valence band of Gr. We recorded high resolution C1s XPS of the Gr/Cu sample at the synchrotron beamline (Stanford Synchrotron Radiation Lightsource at SLAC). The Fermi level in the XAS spectra is at 284.4 eV as obtained from the peak of the C1s XPS spectrum. [18] This XAS spectra is nearly identical to that of graphite which has been much discussed in the literature. [19–22] The central $\pi^*$ peak is thought to be highly excitonic in nature with the electron excited to the Gr valence band localized on the same C with the 1s hole. The region at energies higher to the central peak is thought to arise when the excited electron in the π* valence band is not excitonic and therefore not localized on the C with the 1s hole. [19] There are also higher energy σ* resonances that could not be accessed due to the limited energy range in the PAL experiment.

The changes in the XAS spectrum induced by the optical laser (ΔXAS) at several fixed delay times between the X-ray pulse and the optical laser pulse are shown in the lower part of Fig. 2a. The spectra are obtained by repetitive laser on - laser off measurements at each X-ray energy. Longer temporal delays are given in Fig. S1 for Gr/Cu which shows that the XAS nearly fully recovers to its original spectra over some ~ 1000 ps. A two-dimensional (2D) representation of ΔXAS from similar experiments but taken by scanning the delay at modest time resolution and several fixed energies is given in Fig. 3.

The ΔXAS show several distinct short-lived changes plus some longer lasting ones. There are transient increases in ΔXAS at 282.8 eV and 286.4 eV, a transient decrease in ΔXAS at 284.9 eV and a long-lived decrease in ΔXAS at 285.7 eV near the XAS peak. The temporal response of the principal features of the ΔXAS were obtained by high temporal resolution scans at the fixed energies at or near these features and is shown in Fig. 4.

In all cases the rise times of the features is likely determined by temporal resolution of the experiments, ~ 150 fs. However, the decay times observed are certainly longer than the experimental temporal resolution. In these plots, we account for changes in XAS background, X-ray-induced non-resonant electron emission from the copper substrate, caused by the optical laser, as a step down following the optical laser and with a 100 ps recovery time. Although the decays are likely quite complex, we fit double exponentials as ~85 fs and 1 ps at 282.8 eV; 100 fs and 0.5 ps at 284.9 eV; and a single exponential of 350 fs at 285.7 and 1.1 ps at 286.4 eV. However, since the energies of these features may overlap several physical processes described below (plus the XAS background changes from the optical laser), we only consider the temporal results qualitatively.

Our qualitative interpretation of these results is that the transient increase in ΔXAS intensity at 282.8 eV is created by the holes produced by optical excitation in the formerly fully occupied Gr π band which then allows a C 1s → π transition. The transient decrease in ΔXAS intensity at 284.9 eV results dominantly from the increased population in the π* band created by the optical excitation which lowers the C 1s → π* intensity. The transient increase in ΔXAS at 286.4 eV can arise from two sources; transfer of the initial hot electrons to SCOPS modes and a manybody effect caused by the increased population of delocalized electrons in the π* band. The longer temporal decrease in ΔXAS at 285.7 eV is caused by excitation of both the SCOPS and their subsequent relaxation into the acoustic modes of Gr. Below, we discuss in more detail the basis for these interpretations.

Our CVD prepared Gr/Cu sample is anticipated to be negatively doped so that the Fermi level is ~ 0.4 eV above the Dirac cone of Gr, [23] although the details of the doping can vary with synthesis conditions in the CVD. For the conditions of our experiment (thermal annealing to 300 ºC in UHV), a small gap and n-type doping has been observed. [24] When the Gr/Cu sample is irradiated by the intense optical laser pulse used in the experiments, the conventional two temperature model (2T) for electrons and phonons in the copper substrate implies that the electrons are rapidly heated to $T_e$ ~ 10000 K, followed by a relatively slow thermalization with the lattice modes over several ps to an equilibrium temperature of $T$ ~ 800 K. The 2T model for the conditions of our experiments is shown in Fig. S3. The hot electrons and holes in the Cu can excite the Gr valence bands via charge transfer processes. In addition, direct optical excitation of the Gr overlayer can also occur and is usually described in terms of the 3T model. For the conditions of our experiments the 3T model also predicts high direct excitation of carriers in Gr, but the rapid excitation of the SCOPS limits the electron temperature rise in our observable time regime to ~ 6000 K (Fig. S4).

Because most of the optical energy is deposited into the Cu substrate, we believe that the dominant electronic excitation of Gr occurs from indirect charge transfer processes from the hot electrons and holes created in the Cu substrate. Evidence for this is that the holes created in the Gr π band peak ~ 1.7 eV below the Fermi level in the ΔXAS and hot electrons peak ~ 0.5 eV above the Fermi level in the ΔXAS. If direct optical excitation of Gr was dominate, then the holes created in the π band and the electrons exited in the π* band should follow Fermi-Dirac distributions weighted by the density of the band states, *i e.* both hot holes and electrons peaking close to the Fermi level in the ΔXAS, and this is what is observed for direct optical pumping of graphite [15] and ab-initio theoretical simulations of ΔXAS for a pure Gr film at $T_e$ = 6000 K as shown in Figure 2b. The ab initio calculations for XAS and ΔXAS of a Gr film are described in the Supplementary Material. On the other hand, at 10000 K and even 6000 K the hot holes in Cu are dominated by the d-band even though it is ~ 2 eV below the Fermi energy ($E_F$) while hot electrons peak in the sp-band near $E_F$. This is shown in Fig. S5. However, because injection of holes (electrons) from Cu into Gr occurs through the dipole field created by the n-type doping of Gr, the holes created in Gr will be created at a slightly lower energy than 2 eV relative to $E_F$. In a similar manner, the hot electrons in Gr will be created at slightly higher energies relative to $E_F$. Both shifts are consistent with the hot hole and hot electron peaks observed in the ΔXAS as arising from Cu. This suggests that the electronic coupling between Gr and Cu is so strong that after initial excitation both the Gr and Cu excited states evolve together temporally and are limited by the excitation lifetime in the Cu.

While the excitation and decay of hot carriers in Gr/Cu qualitatively rationalizes the transient changes in the XAS observed at 282.8 eV and 284.9 eV (Fig. 2b), it does not explain the other changes observed in the XAS. The cooling of the hot carriers in free standing or semiconductor supported Gr and graphite via intraband scattering to the SCOPS and then subsequent decay of the SCOPS into the acoustic modes has been well studied. [8–12] In order to understand if these same processes qualitative rationalize the remaining spectral changes observed in XAS, we have made ab-initio theoretical calculations of the XAS for a pure Gr layer that simulate both short-time and long-time behavior that also include excitation of the phonon modes in Gr.

For the short time behavior, we assume that rapid cooling of the hot carriers to 6000 K occurs and that we also populate the SCOPS $(A_1' + E_{2g})$ modes near the $\Gamma$ and K points respectively, in the Brillouin zone to 6000 K (see 3T model in Fig S4). For the long-time behavior, we assume that the Gr cools to thermal equilibrium at 800 K and average the XAS for 50 snapshots of the instantaneous structure taken from an ab-initio molecular dynamics simulation. Details of both calculations are given in the SI. The results of both calculations are also shown in Fig. 2b which compares to the experimental results for 0.2 ps and 6.2 ps respectively. In addition to the transient changes due to hot carrier excitations, we see that excitation of the SCOPS also cause a decrease in the $\Delta$XAS peak height at 285.7 eV as well as an increase in $\Delta$XAS at 286.4 eV caused by a slight blue shift of the peak. Both agree qualitatively with the transient changes in $\Delta$XAS observed in the experiments at 0.2 – 0.5 ps. The calculated $\Delta$XAS via molecular dynamics qualitatively shows that the decrease in the peak intensity and an increase below the peak at ~ 284.7 eV (due to a slight red shift in the peak) observed at 3.2 and 6.2 ps are qualitatively consistent with a longer time thermal distribution of the Gr. Similar changes have been predicted theoretically in other calculations of temperature dependent spectra of graphene [20] and is confirmed by static measurements of the temperature dependence in $\Delta$XAS of our samples via synchrotron XAS measurements (see Fig. S2).

The excitation of SCOPS may not be the only possible mechanism responsible for the feature in the $\Delta$XAS at 286.4 eV since its lifetime is only ~ 1 ps. Since this feature is entirely absent in DFT calculations of hot carrier excitation (see Fig. 2b), we suggest that this could also arise from a manybody effect not included in DFT. The excitation of valence π* electrons in Gr indirectly from hot carriers in Cu should create a delocalized excitation in Gr. It therefore is not anticipated to affect the intensity of the localized excitonic resonance predominately, but rather the higher XAS energy regime in which the electron does not reside on the same C atom as the 1s hole. A similar transient increase in intensity in the high energy side of the XAS peak at ~ 286.4 eV was also observed in optically pumped graphite. [15] We suspect that the same two mechanisms suggested for Gr/Cu could be responsible for this feature in this case as well since optical pumping of graphite will also create SCOPS and create excitations in the delocalized valence band.

Although a theoretical study of the fully coupled electron dynamics of Gr/Cu is beyond the scope of this paper, comparison of the 2T and 3T models in Figs. S3 and S4 which are based on uncoupled components suggest that in the short observable time scales (< 500 fs), that $T_e$ of Cu > $T_e$ of Gr and that this defines the principal direction of electronic excitation as discussed above. After the indirect electronic excitation of Gr, rapid coupling to the SCOPS occurs, followed by a slower decay of the SCOPS into the acoustic modes of Gr. This is a quite similar discussion to the many optical pumping studies of Gr on semiconductors and insulators [8–11] and could account for the observed bi-exponential dependence of the hot hole and electron features at 282.8 eV and 284.9 eV respectively. However, for Gr/Cu the strong electronic coupling of Gr to Cu may mean that the $T_e$ of Gr and Cu evolve together so that the time dependence of the hot hole and hot electrons in Gr reflect the non-exponential lifetimes of the electronic excitations in Cu. We anticipate that there is little direct coupling of the phonons of

Gr with phonon modes in Cu in these short time regimes because of the large interplanar distance of Gr to Cu of ~ 3.3 Angstroms. [23] However, over periods of ~ 1000 ps it appears that the Gr thermalizes with the Cu and the entire Gr/Cu cools to its ambient since the XAS recovers to that prior to the optical laser (see Fig. S1). Time domain transient reflectance experiments on Gr/Cu have been interpreted to suggest strong phonon coupling between Gr and Cu in the ps regime, but they neglect charge transfer and its effects on the reflectivity in the short time regime. [25]

In this paper we have shown that fs optical pumping and X-ray absorption spectroscopy at the PAL free electron laser can directly probe the electron dynamics of a graphene monolayer adsorbed on Cu in the femtosecond regime. By analyzing the results with ab-initio theory we infer that the excitation of graphene is dominated by indirect excitation from hot electron-hole pairs created in the Cu by the optical pules. However, once the excitation is created in graphene, its decay follows a similar path as in many previous studies of graphene adsorbed on semiconductors, *i e.* rapid excitation of SCOPS and eventual thermalization. It is likely that the lifetime of Cu hot electron-hole pairs governs the lifetime of the electronic excitation of the adsorbed graphene.

**Acknowledgements:**
This research was supported by the U.S. Department of Energy, Office of Science, Office of Basic Energy Sciences, Chemical Sciences, Geosciences, and Biosciences Division, Catalysis Science Program to the Ultrafast Catalysis FWP 100435 at SLAC National Accelerator Laboratory under Contract DE-AC02-76SF00515.  This study was also partly supported by the National Research Foundation (NRF-2020R1A2C1007416) of South Korea. We gratefully acknowledge the technical support by the XFEL accelerator and beamline divisions at Pohang Accelerator Laboratory, and the Stanford Synchrotron Radiation Lightsource, SLAC National Accelerator Laboratory, supported by the U.S. Department of Energy, Office of Science, Office of Basic Energy Sciences under Contract No. DE-AC02-76SF00515.

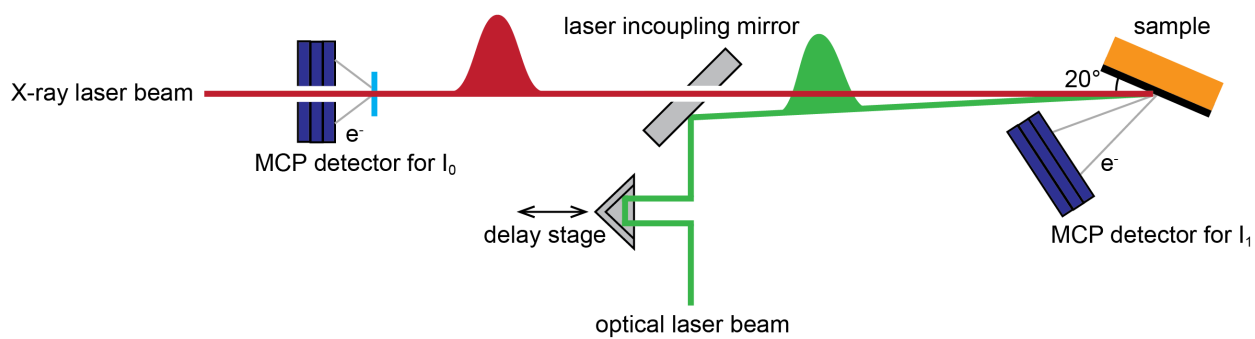

Figure 1. Schematic of the experimental optical pump - X-ray spectroscopy probe for measuring femtosecond electron dynamics of graphene adsorbed on Copper. The SSS beamline equips a collinear in-coupling geometry of the optical laser and X-ray beams, which maximize time resolution and ease of spatial alignment. An MCP (microchannel plate) detector for $I_0$ and $I_1$ respectively monitors the intensity of incoming X-ray beam and absorbed X-ray by the sample.

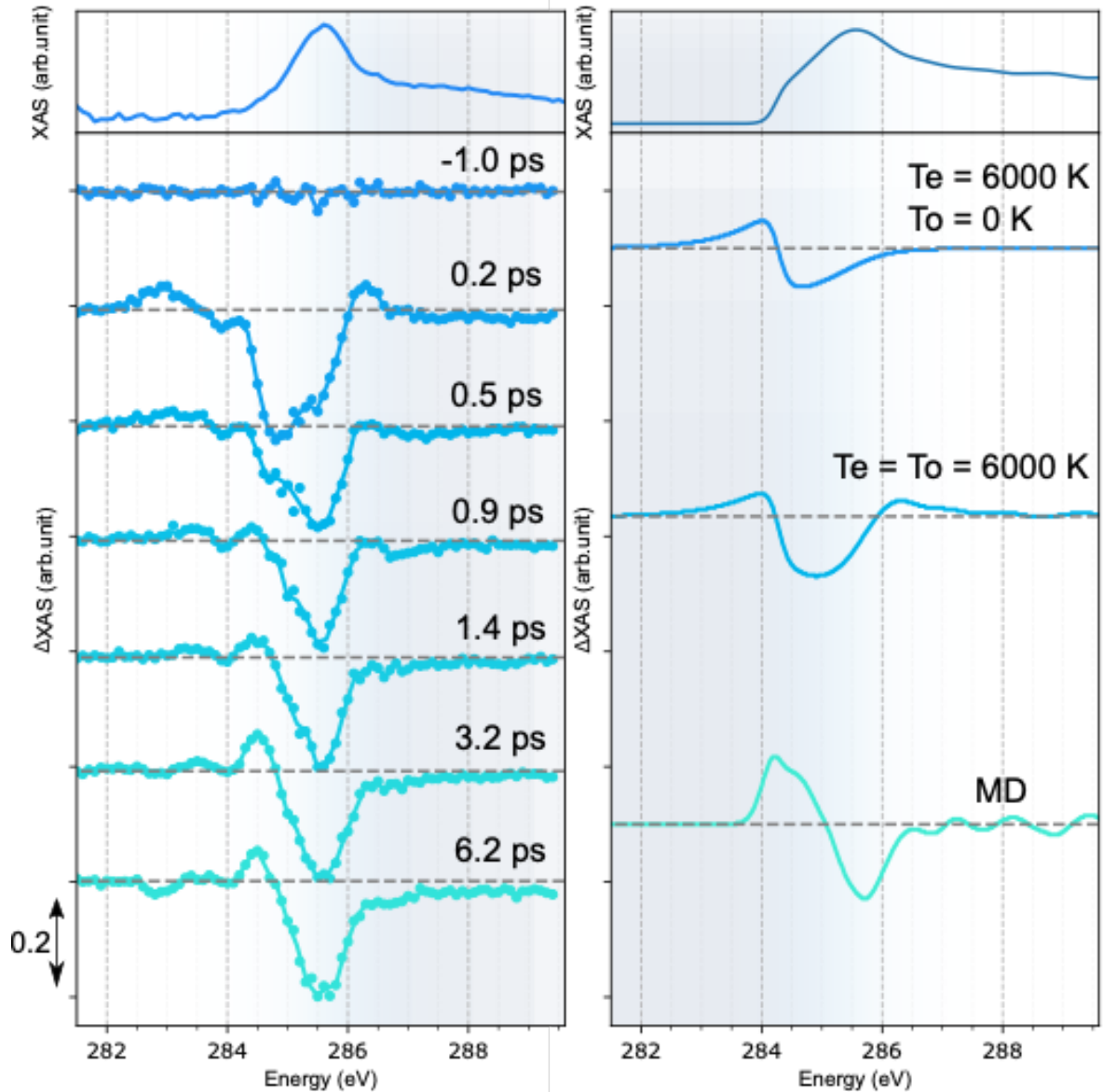

Figure 2. (a) XAS spectra of Gr/Cu (top) and $\Delta$XAS spectra of Gr/Cu for the labeled delay times between the optical pump and X-ray probe lasers. The negative delay probes the unexcited sample. (b) Theoretical XAS and $\Delta$XAS of a Gr layer for the cases labeled that represent various stages of Gr temporal evolution; $T_e$ =6000 K represents the shortest delay times probed in the experiment) when only thermalized electronic excitation occurs, $T_e = T_o$ = 6000 K represents the shortest delay times probed with both electronic and SCOPS excitation and MD qualitatively represents the longest delay times probed.

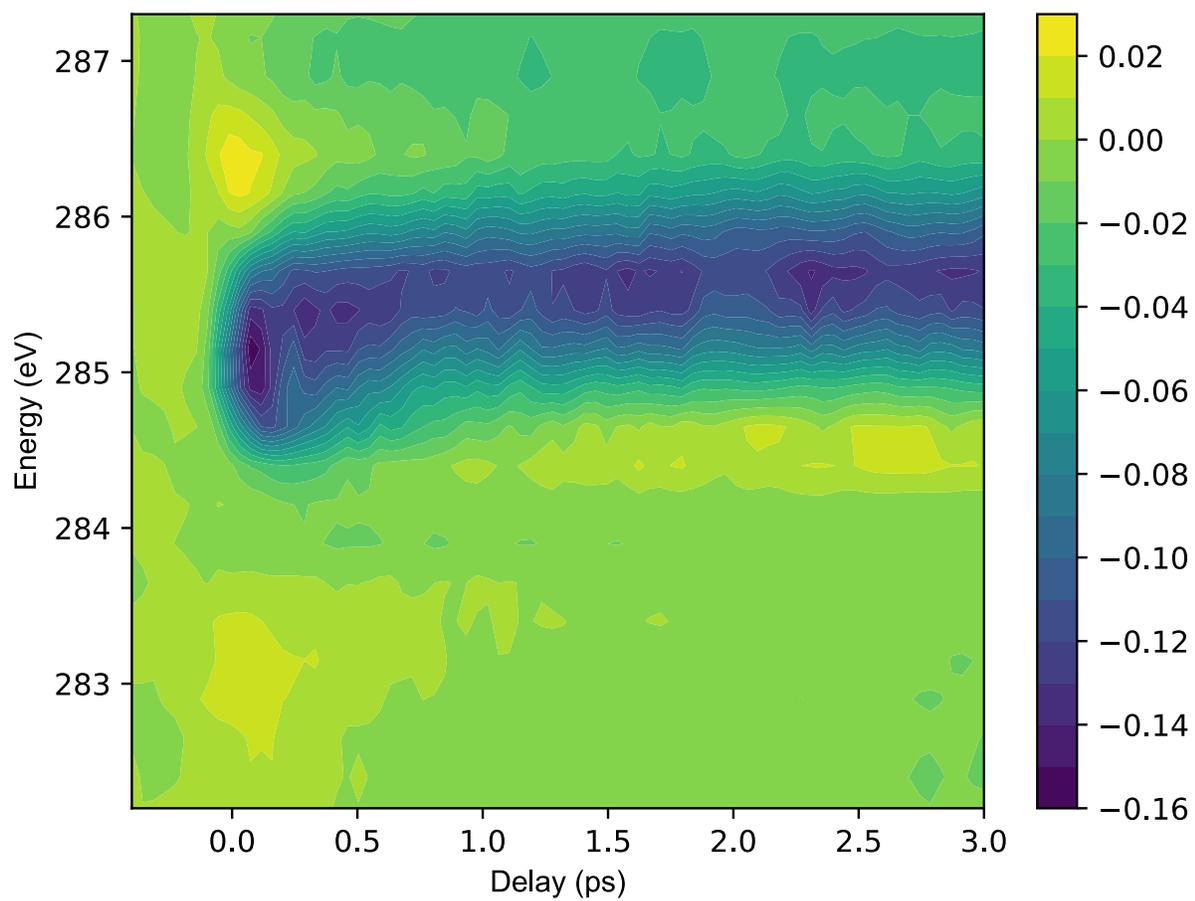

Figure 3. Two dimensional representation of ΔXAS of Gr/Cu in terms of X-ray energy and delay relative to the optical pulse. The heat map is given to the side.

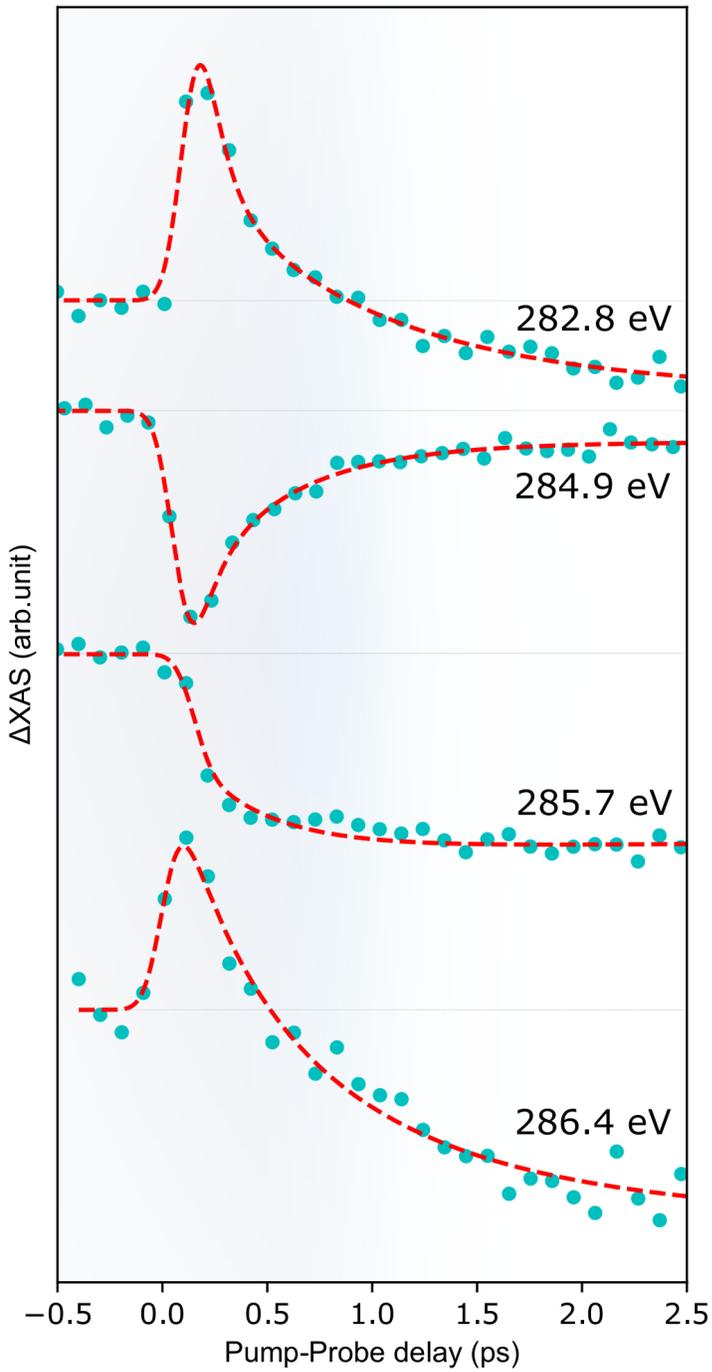

Figure 4. Temporal response of ΔXAS of Gr/Cu at the principal features where changes occur in the ΔXAS. The intensity of each scan was normalized by the absolute value at the peak maximum. The curves represent bi-exponential or single exponential fits to the data plus a step background change in Cu photoelectrons caused by the optical laser.

# Supplementary Material

## 1. Experimental details:

### 1.1 Sample preparation

The Gr/Cu sample was obtained from Grolltex, Inc. and was prepared by CVD growth of graphene on a 25 micron-thick Cu foil. The sample quality was screened on SSRL BL8-2 using angle- dependent drain current detection XAS. Prior to the measurement in the analysis chamber, the sample was treated by in-situ heating in the preparation chamber under a high vacuum to remove adsorbed contaminants. In the heating treatment, the samples were first outgassed at the temperature of 150 °C for 15 minutes, followed by 15 minutes stepwise increase in temperature from 50 °C to 300 °C. Subsequently, the sample was kept at 300 °C for more than an hour. After cooling to room temperature, the sample was transferred from the preparation chamber under high vacuum conditions to the analysis chamber also under high vacuum for time- resolved XAS measurements.

### 1.2 Time-resolved XAS measurements

We used the SSS beamline at PAL [1,2], which contains the surface-science sample preparation facilities, timing tools, as well as the partial electron detector. The experiments were performed in an optical pump/X-ray probe scheme using the optical-laser pulses provided by the PAL XFEL optical-laser system for triggering the surface excitation. For this experiment, 100 fs-duration, 400 nm center-wavelength pulses ranging up to 100 µJ came from its frequency-doubled output.

The pump pulse fluence was controlled to avoid observable optical damage to the graphene overlayer on the sample. We ran the experiment at 75% of the damage threshold fluence, which was sufficient to leave the C XAS spectrum stable throughout the entire time of many measurements. This was also approximately 50% of the physical damage threshold of the Cu substrate. The DXAS results at lower optical fluences were identical to that observed at the higher fluence except for their smaller amplitudes.

The PAL XFEL ultrashort soft X-rays, with a pulse width of <50 fs and a photon energy around the C K-edge at ~270-290 eV, probed the resulting excitation of the Gr adlayer. Since the laser and X- ray pulse repetition frequencies were set respectively to 30 and 60 (i e. 2 x 30) Hz, the laser pumping was in effect "chopped," enabling synchronous detection of slight changes in the Gr adlayer's X-ray absorption. As shown in Figure 1 of the text, the laser and X-ray beams propagated nearly collinearly. This arrangement simplified alignment (spatial overlap,) reduced spot walk off as the optical delay was scanned and minimized *de facto* pulse blurring/stretching (that would result from oblique incidence.) At the sample position, the optical laser beam formed a Gaussian profile (diameter of 200 µm at the normal incidence). The comparatively smaller X-ray spot (70 µm at normal incidence) could be entirely covered by the optical laser-pumped zone, so changes in excitation efficiency due to spatial-overlap shifts were not a significant concern. The relative timing between the pump and probe pulses was controlled with a mechanical delay stage in the laser beam. Spatial and temporal ("time zero") overlap was accomplished with the aid of a thin Ce:YAG crystal whose blue-laser-induced-fluorescence quantum yield was altered by X-ray illumination. The fluctuation in excitation efficiency due to the drift in the spatial overlap between X-ray and optical laser was minimized by using an X-ray spot size (70 µm at the normal incidence) smaller than the size of the pseudo-flat top area of the optical pump laser.

### 1.3 Auger detection of Graphene

For pump-probe measurements, X-ray and optical laser beams impinging on the sample surface at a near-grazing ~20° incidence angle to the sample surface. We used monochromatic X-rays, $\Delta E \leq$ ~200 meV, p-polarized relative to the surface of Gr/Cu. Absorption of X-rays by core-electrons results in the emission of Auger electrons from the sample. Consequently, the carbon KVV Auger electron yield is proportional to the absorption coefficient. A biased microchannel plate (MCP) detector was employed to measure the carbon KVV Auger electron yield from the sample ($I_1$ signal). A retardation voltage of -60 V was applied between the sample and the MCP input electrode to reject low kinetic energy electrons photogenerated by the pump optical-laser pulse, and also partially discriminates the low kinetic energy secondary electron from the primary high- energy C KVV Auger electron. The MCP for the $I_1$ signal can also detect X-ray induced photoelectrons from the copper M-shell and d-band (and minorly from the C & Cu sp valence states), which occur at equal or higher kinetic energy than that of carbon KVV Auger electrons. Notably, the ratio between the C KVV Auger yield and the Cu photoelectron yield in the $I_1$ signal changes as a function of photon energy. The $I_1$ signal measured below the absorption threshold solely represents the photoelectron yield of the copper M-shell and Cu and C valence states. At the central π* peak, the contribution of these photoelectron yields to the $I_1$ signal is ~ 50%.

### 1.4 Absorbed Optical Energy

We estimate the absorbed pump fluence in Cu in the probed region, the pseudo-flat-top area, based on the sample's reflectivity and absorption coefficient using a simple Fresnel analysis for the experimental conditions and ignoring the small absorption in the Gr overlayer. This predicts ~ 67% absorption of the optical flux in Cu. The direct optical absorption in the Gr adlayer was estimated to be ~ 2 % for our experimental conditions.

### 1.5 Time-resolved XAS data analysis

X-ray absorption spectra (XAS) were obtained by plotting the $I_1$ signal as a function of the X-ray energy. For normalization, an MCP-detector incoming-photon-flux monitor upstream of the Kirkpatrick–Baez (KB) mirror system measured the total number of secondary electrons produced at the beam-exit window coated with platinum ($I_0$ signal). Carbon in the boron carbide-coated KB mirror system imposed an energy dependence on the X-ray-beam transmission efficiency. To account for this, we recorded the X-ray-energy-dependent photoionization yield of an Argon gas jet placed after the KB mirror system. The ~290.7 eV $\pi^*$ peak positions of gas-phase $CO_2$ [3] were used to calibrate the energy scale of the PAL X-ray monochromator.

A differential technique increased the detection sensitivity to small differences between secondary-electron signals collected in the pump-laser-on and pump-laser-off conditions. The $I_0$ and $I_1$ signals were alternately measured in the 30 Hz pump-laser-on/pump-laser-off sequence using a high-speed digitizer with a coincidence window of 14.2 ns. Before subtracting the laser-on and laser-off $I_1$ signals, the offset in the $I_1$ signal was corrected. In energy scan measurements at a fixed delay, the laser-on $I_1$ and laser-off $I_1$ signals were scaled by the number of probed copper atoms, which is estimated from the intensity of the $I_1$ signal at an X-ray energy below the C K-edge absorption threshold. In the delay scan measurements at fixed energies, there can be a time-dependent change in the Cu d-band occupation, or a thermal expansion of the Cu lattice caused by the optical laser. These reduce the photoelectron yield in the $I_1$ signal and appear as a time-dependent offset in the data.

For the delay scans at fixed X-ray energy, we consider a negative time delay (X-rays precede 400 nm pump), where the electronic structure of carbon is identical between the laser-on and laser-off conditions. Since the amount of photoionized atoms is the same, the offset was adjusted to align the laser-on $I_1$ signal in the delay scans to have the same intensity as that of the pump-laser- off $I_1$ signal at this energy. Twenty-one delay scans at different x-ray energies were assembled to construct a 2D representation of the time-resolved difference XAS (Fig. 3 in the text).

### 1.6 Static XAS measurement at SSRL

Static XAS was measured at SSRL BL13-2 using Auger electron yield as a function of incoming X-ray photon energy. An electron spectrometer, Gammadata-Scienta SES-100, measured the intensity of 240-270 eV carbon KLL Auger electrons ejected from the sample. Secondary electrons from the copper substrate also appear in the same energy window contributing a background to the measured signal. We measured a background spectrum separately by monitoring the intensity of 130-150 eV secondary electron, which dominantly comes from the copper substrate, as a function of incoming X-ray photon energy. The carbon KLL Auger electron yield and Cu background yield spectra were normalized, assuming that both detect the electron emission from the copper substrate at X-ray photon energies below the C K-edge absorption threshold. Given this analysis, we were able to measure the temperature dependence of the Gr/Cu XAS (see Fig. S2). The he $\pi^*$ peak positions of gas-phase $CO_2$ at 290.7 eV [3] were used to calibrate the energy scale of the SSRL monochromator.

### 2. Additional Experimental results

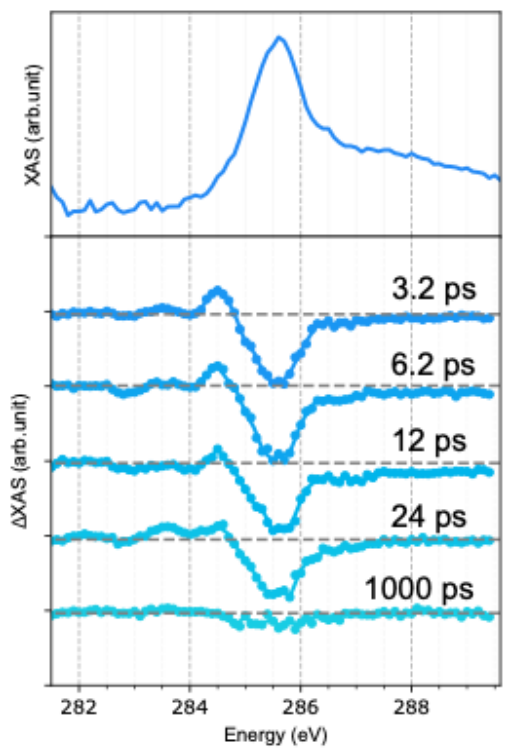

**Figure S1.** Long time response in ΔXAS of a sample of Gr/Cu at the delay times listed.

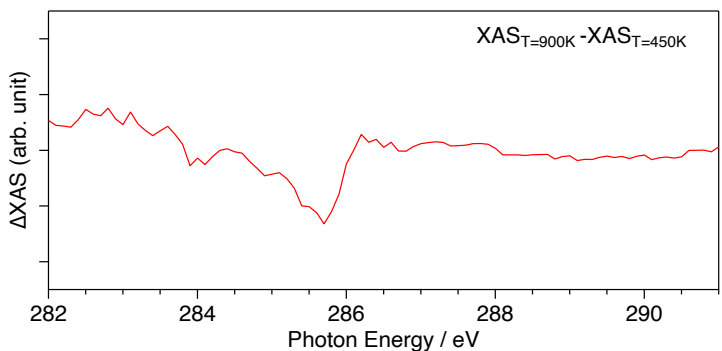

**Figure S2**. A spectral difference between XAS of Gr/Cu taken at T=900 K and T=450 K at SSRL

## 3. Theoretical Details

### 3.1 Structure relaxation and XAS calculation of Graphene layer

Structural relaxation, molecular dynamics (MD) and phonon potential energy surface calculations were performed with the VASP code [5] and projector-augmented wave frozen-core pseudopotentials [6]. For electronic exchange and correlation, the PBE functional [4] was employed. For the plane-wave basis sets, a 500 eV cutoff was used. XAS simulations were performed with the XSpectra code [7,8] from the plane-wave basis pseudopotential DFT package Quantum Espresso [9]. Core excitations were modeled with frozen core-hole pseudopotentials. A carbon pseudopotential with half a core-hole in the 1s shell was employed according to the transition potential approach to the calculation of dipole matrix elements [10]. XAS onset shifts due to changes in atomic structure were computed as total energy differences of carbon full core-hole and ground state pseudopotential-based DFT simulations. For details of the carbon core-hole pseudopotentials and their benchmarked performance we refer to Ref. [11]. The MD simulations were performed with a $4\sqrt{3} \times 4\sqrt{3} \times 1$ supercell of graphene, with a $3 \times 3 \times 1$ k-points mesh sampling the Brillouin zone. Brillouin-zone sampling for all other simulations was performed with k-point spacings of at most 0.07 Å$^{-1}$. MD simulations were performed at $T$=300 and 2000 K respectively for 10 ps with a time step of 0.1 fs. MD snapshots at every 100 fs were used as structural input for the XAS simulations, and the resulting spectra were averaged.

For the SCOPs $E_{2g}$ at Γ and $A1'$ at K, frozen-phonon calculations were performed with VASP on free-standing graphene to determine the potential energy surfaces with these modes including anharmonic effects. For the E2g modes, two high-symmetry directions were considered, lowering the symmetry of graphene from *P6/mmm* to *P2/m* and *Cmmm*, respectively. For the A1' modes, two high-symmetry directions leading to *P-6m2* and *P6/mmm* symmetry (with six carbon atoms per primitive unit instead of two at Γ), respectively, were considered. Along these modes, 50 displacements with a maximum amplitude of 0.3 Å were considered. The corresponding 1D quantum anharmonic oscillator problem for each mode was solved numerically. The probability densities of the oscillator states were used to weight XAS spectra simulated along the 50 snapshots of the frozen phonon paths. The resulting contributions from the lowest energetic oscillator states with quantum number *n* up to 9 were weighted with a Boltzmann distribution.

### 3.2 Two temperature model of Copper

A two-temperature model [12] with electron temperature $T_e$ and phonon or copper lattice temperature $T_{ph}$ is solved numerically to determine the evolution of these temperatures as a function of time $t$ after pump laser excitation and penetration depth $z$ into the copper surface

$$C_e(t,z) \frac{\partial T_e(t,z)}{\partial t} = \frac{\partial}{\partial z}\left(\kappa(t,z) \frac{\partial T_e(t,z)}{\partial z}\right) - g \cdot [T_e(t,z) - T_{ph}(t,z)] + S(t,z)$$

$$C_{ph}(t,z) \frac{\partial T_{ph}(t,z)}{\partial t} = g \cdot [T_e(t,z) - T_{ph}(t,z)]$$

$$S(t,z) = I(t) \frac{\exp(-z/\lambda)}{\lambda} .$$

$C_e = \gamma T_e(t,z)$ is the electronic heat capacity with specific heat $\gamma$. $\kappa(t,z) = \kappa_0 T_e/T_{ph}$ is the thermal conductivity and $\kappa_0$ the electronic heat conductivity. $g$ is the electron-phonon coupling. The phonon heat capacity $C_{ph}(t,z)$ is computed by numerically integrating a Debye model with Debye temperature $\Theta_D$ and atomic density $n$. $I(t) = F \exp\left(-\frac{t}{2\sigma^2}\right)/\sqrt{2\pi\sigma^2}$ is the assumed profile of the pump laser pulse. $F$ is the absorbed laser fluence. $\sigma = \Gamma/2.355$, where the pulse duration $\Gamma$ is taken at full width at half maximum. $\lambda$ is the optical penetration depth. The constants $\gamma = 98$ Jm$^{-3}$K$^{-2}$, $\kappa_0 = 401$ Wm$^{-1}$K$^{-1}$, $g = 10^{17}$ Wm$^{-3}$K$^{-1}$, $\Theta_D = 343$ K, $n = 8.5 \cdot 10^{28}$ m$^{-3}$ and $\lambda = 14.4$ nm for copper were taken from Ref. [13]. The experimental parameters are $F = 220$ Jm$^{-2}$ and $\Gamma = 50$ fs.

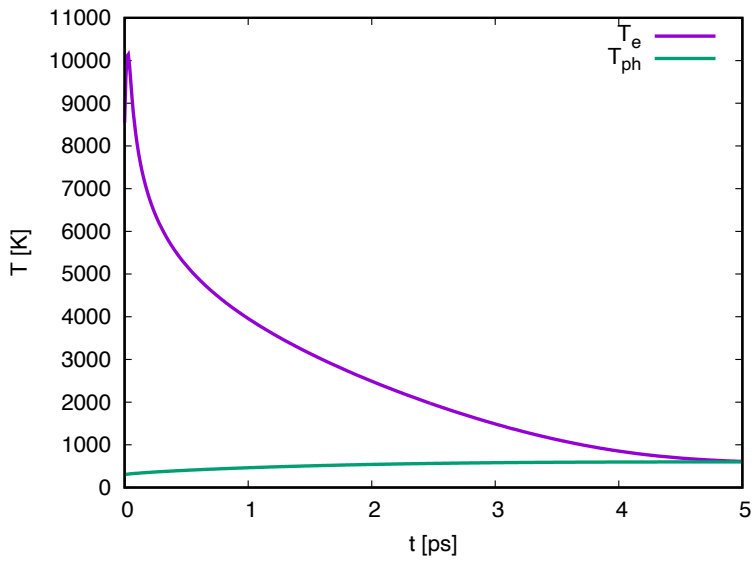

**Figure S3.** Two temperature model for Cu excitation by the optical laser at the PAL experimental conditions. $T_e$ is the electron temperature and $T_{ph}$ is the phonon temperature.

### 3.3 Three temperature model for graphene

The three-temperature model was revised from the model reported in Ref. [14]. The electronic temperature $T_{el}(t)$ and $T_{op}(t)$ satisfies the following two equations.

$$\frac{dT_{el}(t)}{dt} = \frac{I(t) - \Gamma(T_{el}, T_{op})}{c_e(T_{el})}$$

$$\frac{dT_{op}(t)}{dt} = \frac{\Gamma(T_{el}, T_{op})}{c_{op}(T_{op})} - \frac{T_{op}(t) - T_0}{\tau_{op}}$$

where

$$\Gamma(T_{el}, T_{op}) = \beta\{[1 + n(T_{op})]\int_{-\infty}^{+\infty} D(E)D(E - \hbar\Omega)f(E, T_{el})[1 - f(E - \hbar\Omega, T_{el})]dE$$
$$- n(T_{op})\int_{-\infty}^{+\infty} D(E)D(E + \hbar\Omega)f(E, T_{el})[1 - f(E + \hbar\Omega, T_{el})]dE\}$$

$c_{op}(T_{op}) = -4.79 \times 10^9 + 1.82 \times 10^7 T_{op} + 1.34 \times 10^4 T_{op}^2 + 5.16 T_{op}^3$ (in units of eVcm$^{-2}$K$^{-1}$)

According to the Dulong–Petit law, the Cv should be smaller 24.94 J/mol/K. So $c\_op$ ($T_{op}$) was capped to 24.94 J/mol/K whenever the calculated value is larger than it.

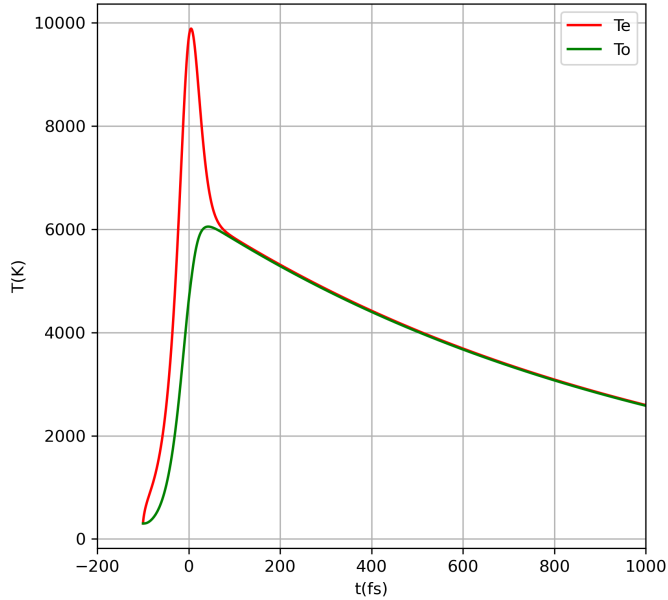

**Figure S4.** Three temperature model for Gr under the conditions of the PAL experiment. T$_e$ is the electron temperature and T$_o$ is the SCOPS temperature which decays over ~ 1 ps into acoustic modes.

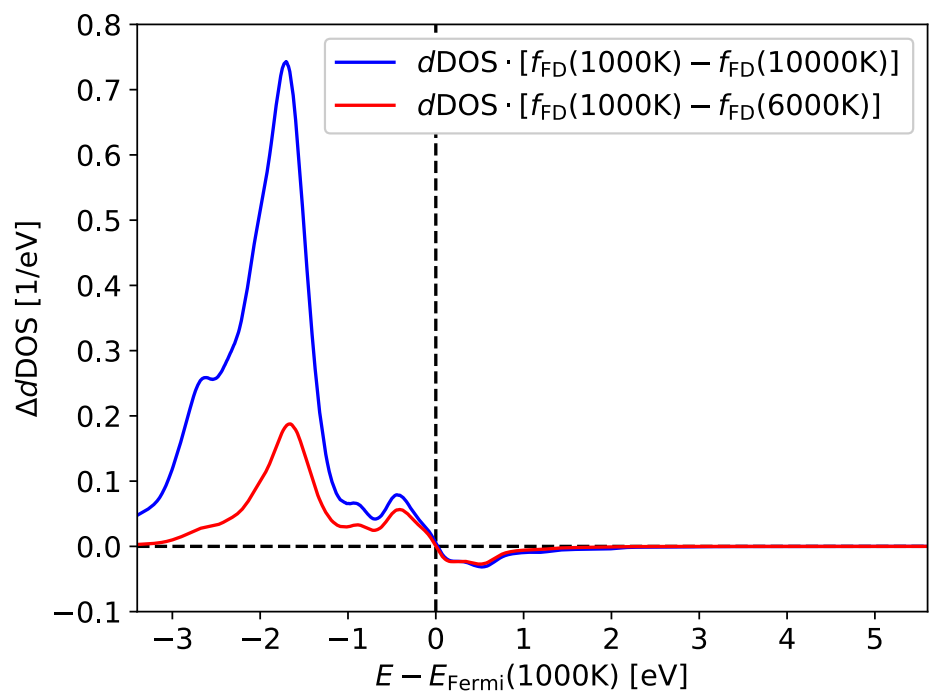

**Figure S5.** Plot of Cu DOS projected onto d-states of a surface Cu(100) atom (in a 6-layer Cu slab) at 1000 K (assumed equal to that at 300 K) relative to that at 10000 K and 6000 K. In this representation, holes are positive and electrons are negative.